\documentclass[%
prl,
superscriptaddress,
preprint,
 amsmath,amssymb,
 aps,
]{revtex4-2}

\usepackage[dvipsnames]{xcolor}

\usepackage{graphicx}
\usepackage{dcolumn}
\usepackage{bm}
\usepackage{todonotes}
\usepackage{caption}
\usepackage{hyperref}

\hypersetup{
    colorlinks,
    linkcolor={blue},
    citecolor={blue},
    urlcolor={blue}
}

\begin{document}

\preprint{PRL Draft}

\title{Identifying the structure of dynamical transitions in logistic map}

\author{Aswin Balaji}
 \altaffiliation[Current affiliation: ]{Department of Physics \& Astronomy, Northwestern University, Evanston, Illinois 60208, USA} 
 \affiliation{%
 Department of Aerospace Engineering, Indian Institute of Technology Madras, Chennai 600 036, India
} 
\affiliation{%
 Centre of Excellence for Studying Critical Transitions in Complex Systems, Indian Institute of Technology Madras, Chennai 600 036, India
}%
\author{Shruti Tandon}%
 \email{shrutitandon97@gmail.com}
\affiliation{%
 Department of Aerospace Engineering, Indian Institute of Technology Madras, Chennai 600 036, India
}%
\affiliation{%
 Centre of Excellence for Studying Critical Transitions in Complex Systems, Indian Institute of Technology Madras, Chennai 600 036, India
}%
\author{Shwetha Viswesh}%
\affiliation{%
 Department of Mechanical Engineering, Indian Institute of Technology Madras, Chennai 600 036, India
}%
\affiliation{%
 Centre of Excellence for Studying Critical Transitions in Complex Systems, Indian Institute of Technology Madras, Chennai 600 036, India
}%
\author{Norbert Marwan} 
\affiliation{%
 Potsdam Institute for Climate Impact Research (PIK), Member of the Leibniz Association, Potsdam 14473, Germany
}%

\affiliation{%
 Institute of Geosciences, University of Potsdam, Potsdam 14476, Germany
}%

\affiliation{Institute of Physics and Astronomy, University of Potsdam, Potsdam, 14476, Germany}

\author{J{\"u}rgen Kurths}
\affiliation{%
 Potsdam Institute for Climate Impact Research (PIK), Member of the Leibniz Association, Potsdam 14473, Germany
}%
\affiliation{%
 Institute of Physics, Humboldt Universit\"at zu Berlin, Berlin 12489, Germany
}%

\author{R. I. Sujith}

\affiliation{%
 Department of Aerospace Engineering, Indian Institute of Technology Madras, Chennai 600 036, India
}%
\affiliation{%
 Centre of Excellence for Studying Critical Transitions in Complex Systems, Indian Institute of Technology Madras, Chennai 600 036, India
}%




\date{\today}

\newpage

\begin{abstract}
\textbf{Abstract:} 
Nonlinear dynamical systems manifest rich variety of dynamical states and transitions driven by fluctuations. To understand the pattern of fluctuations during dynamical transitions, we investigate the structural features of chaos to order transition in logistic map. We determine fluctuations as amplitude jumps and encode them onto a complex network where nodes represent amplitude levels and links represent transitions between distinct amplitude bins. We discover that global network measures identify points of period doubling, regimes of periodicity and chaos, including interior crises events. Using local network measures, we also unravel novel peculiar $\subset$-shaped patterns in the orbit diagram that we show are reminiscent of the distribution of stable and unstable periodic points in the bifurcation diagram.

\end{abstract}

\maketitle


\noindent Complex systems ranging from neuronal to astronomical scales exhibit a variety of dynamical states and transitions, often with similar features. The nonlinear functional relations within such a system give rise to a wide spectrum of behavior in the temporal and spatiotemporal domains. Oscillatory behavior in physiology \cite{Glass2001SynchronizationPhysiology}, transition from chaos to order via the intermittency route in thermo-fluid systems \cite{Sujith2021ThermoacousticInstability,Nair2014IntermittencyCombustors}, chaotic and mixed-mode oscillations of ions in Belousov-Zhabotinsky reaction \cite{Prigogine2018OrderNature}, abrupt transition in climatic variables \cite{Lockwood2001AbruptReview}, and chaotic fluctuations of ionospheric density \cite{Bhattacharyya1990ChaoticMeasurements} are some of the many examples of dynamical states and transitions observed in real-world systems.
Characterizing the pattern of fluctuations across different scales not only aids in a better understanding of the underlying processes but also helps us develop precursors for critical transitions in complex systems \cite{Balaji2025NonmonotonicSystems}. 
Dynamical transitions can be driven by fluctuations, whereby sufficiently large fluctuations can cause the system to transition from one attractor to another. Consequently, the amplitudes of fluctuations during a specific dynamical regime provide information about the underlying structure of the phase space. For instance, in a phase space with multiple periodic orbits, fluctuations induce transient excursions from around one orbit to another, thus also modulating the amplitude of the dynamics. Here, we investigate the structure of fluctuations during various dynamical states revealing novel intricate patterns of stable and unstable periodic points in a quadratic map. Our approach provides a physics-informed data-driven method to extract the stable and unstable periodic orbits from time series.

Inferring accurate reduced-order models or finding analytical solutions for real-world systems is challenging. Data-driven methods are becoming popular as these approaches help to identify hidden patterns in data and unravel the complexity in real-world dynamical systems \cite{Zanin2016CombiningHowb}. Nonlinear time series analysis tools were developed to characterize invariant characteristics of the phase space, such as generalized fractal dimension and Lyapunov exponents. These methods are however, often limited to low-dimensional systems, are sensitive to parameter selection, and are not reliable in the presence of noise \cite{Marwan2002Recurrence-plot-basedData}. More recently, complexity measures based on symbolic dynamics, such as Rényi entropy \cite{Renyi1961OnInformation} and mutual information \cite{Cover2005EntropyInformation} were proposed as alternative means to characterize system dynamics. Several entropy-based complexity measures have been proposed to capture (at least partly) the complexity of different dynamical states and transitions in various complex systems \cite{Ladyman2013WhatSystem}. 
In this Letter, we investigate the dynamics exhibited by the logistic map ($x_{n+1} = rx_n(1-x_n)$), which displays rich dynamics such as periodicity, chaos, intermittency, and chaos-chaos transition for different values of the control parameter $r$. The logistic map was initially developed to model biological population \cite{May1976SimpleDynamics}. The dynamical states and transitions generated by this paradigmatic map are also observed in a wide variety of natural and experimental systems \cite{Crevier1998SynchronousMan,Jia2011DynamicsPatterns,Sterling1993NonlinearCombustor, Chialvo1990LowTissue}. 

\begin{figure*}
     \centering
     \includegraphics[width=0.9\linewidth]{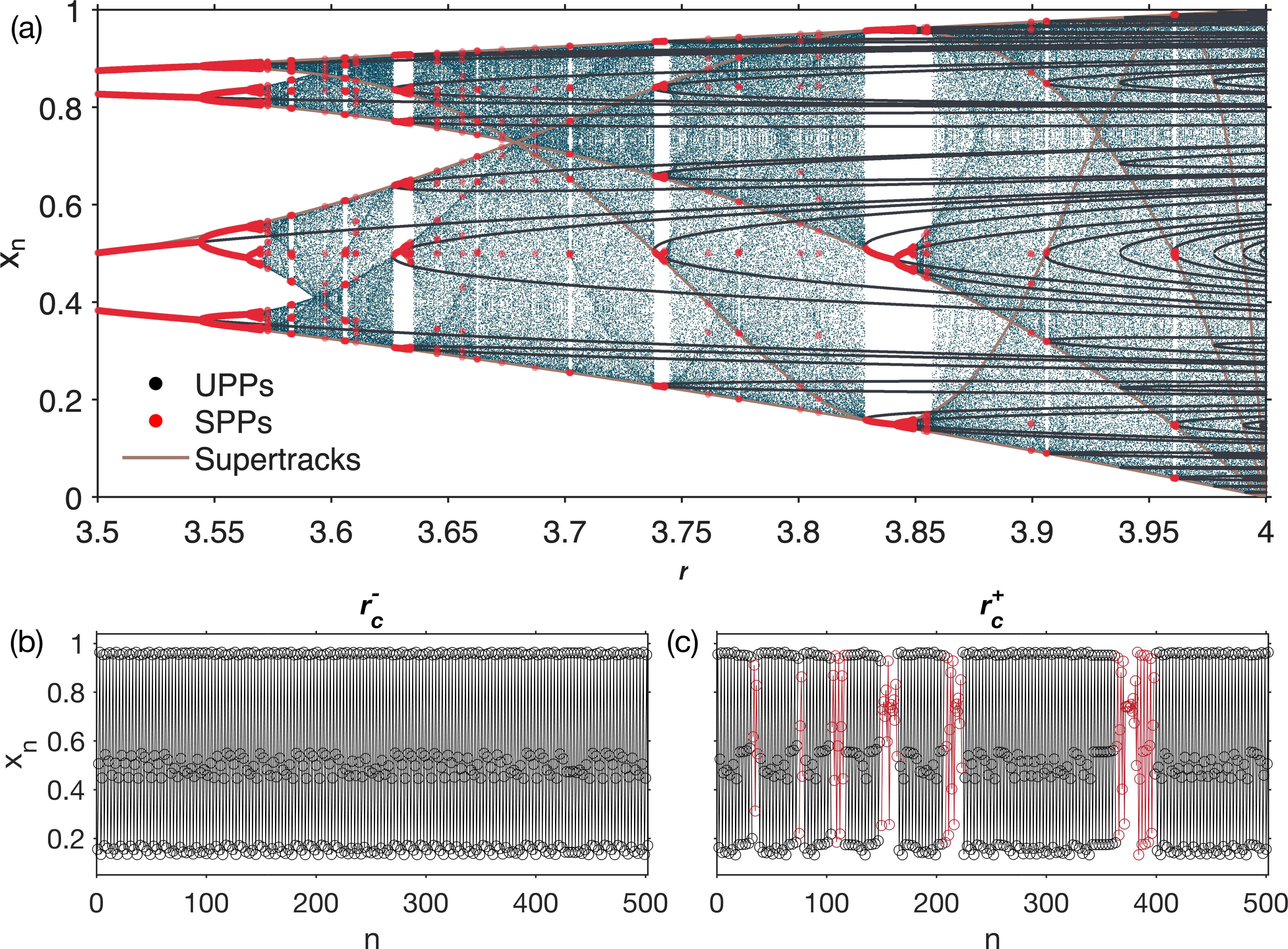}
     \caption{(a) The orbit diagram of logistic map (gray-green points) for $r \in [3.5,4]$, superimposed on the first few orders of supertracks (brown lines), unstable periodic points (UPPs; black lines), and stable periodic points (SPPs; red points). For each value of $r$, trajectories are iterated for 200 steps after discarding 1000 transients to capture the attractor dynamics. (b,c) The sequence generated from the logistic map in the vicinity of $r_c = 3.857\dots$. For $r$ slightly less than $r_c$ ($r_{c}^{-}$ ), the orbit switches predominantly between three chaotic bands. With a slight increase in the control parameter ($r_{c}^+$), the orbit intermittently switches between a wide range of amplitudes of a new broad attractor (red markers) including amplitudes of the attractor at $r_{c}^{-}$. Such intermittent switching during a crisis is referred to as crisis-induced intermittency.}
     \label{fig:orbit_supertrack}
 \end{figure*}

For a given initial condition $x_{0}$, the sequence of solutions \{$x_{0}$, $x_{1}$, $x_{2}$, \dots \} visited by the logistic map at a fixed $r$ is called the orbit or trajectory. The attractor (also known as the orbit diagram) of the system shows the trajectories as a function of the control parameter $r$ (see attractor in dark green in Fig.~\ref{fig:orbit_supertrack}(a)). The map exhibits a period doubling route to chaos with increase in $r$, where $r_{\infty}= 3.569\dots$ is the first point of onset of chaos and is referred to as an \textit{accumulation point}. One would imagine that the map would continue to exhibit chaos after $r_{\infty}$, but interestingly, the system switches back to periodic dynamics for certain range of $r$ (referred to as periodic windows). Specifically, beyond $r = 1 + \sqrt{8}$, the map exhibits cycles of period-3. \textcite{Li2004PeriodChaos} proved that the existence of a stable period-3 cycle in an iterative map implies the existence of a stable cycle of any period for some other control parameter. 

Further, we observe `emergent curves' due to high-density points accumulating in the orbit diagram along these curves as $r$ varies (Fig.~\ref{fig:orbit_supertrack}(a)). The closed-form solution for such curves can be obtained by iterating the map exactly from the critical point $x_c = 0.5$, given by $s_{i+1}(r) = r s_{i}(r) (1 - s_{i}(r)) \text{ where, }  s_{0} = 0.5$, and, $i$ denotes the order of the curve. These curves, called \textit{super track functions}, intersect with the fixed point of the map at $1 - 1/r$. Other intersection points of these curves coincide with different basic features of the map, such as band merging and laminar states \cite{Marwan2002Recurrence-plot-basedData}. Apart from supertrack functions evident in the orbit diagram, the logistic map also depicts other patterns owing to the bifurcation diagram. A bifurcation diagram captures the evolution of the dynamical states across varying $r$, and contains \textit{all} fixed points (stable, unstable and neutral) of the iterated maps $F^p$, known as $P$ curves \cite{ROSS2009TheToo} that reveal the underlying periodic-orbit structure of the map. During chaotic regimes, the tendency of the trajectory to approach a point depends on the relative stability of the associated branch of fixed points. That is, the stability of a point at fixed $r$ depends on the order of bifurcation at which the stable/unstable branch containing this point was created with varying $r$.

To identify these stable and unstable periodic fixed points, we compute the solutions to $F^p(x) =x$, where $F=rx(1-x)$ and $p$ is the period. For each $r$, the roots of the equation are searched on a uniform grid of 2000 points within the the interval $x \in [10^{-6},1-10^{-6}]$ to avoid the endpoints. Exact periodicity is rigorously enforced by excluding any solutions that satisfy $F^p(x) = x$ for integer divisor $q < p$ (up to tolerance $10^{-8}$) and eliminating duplicates by rounding to twelve significant digits. Linear stability of the roots are evaluated using the multiplier $\lambda = \Pi_{k=0}^{p-1} r(1-2x_k)$, where roots yielding $|\lambda| < 1$ are classified as stable periodic points (SPPs), while those with $|\lambda| > 1$ are unstable periodic points (UPPs), as shown in Fig.~\ref{fig:orbit_supertrack}(a). The UPPs and SPPs are obtained with an accuracy up to sixth decimal point after removing transients and extracting the periodic points in long-time dynamics.

Shifting focus to the vicinity of $r = 1 + \sqrt{8}$, we see that the chaotic attractor undergoes a sudden change in size; such sudden destruction or widening of a chaotic attractor is referred as interior crises \cite{Grebogi1983CrisesChaos,Wackerbauer1994AMeasures}, which are also observed in other nonlinear maps and real-world systems \cite{Jeffries1983DirectOscillator,Ditto1989ExperimentalExponent}. Such transitions are accompanied by unique characteristics and often difficult to detect using typical time series analysis. 
\begin{figure*}
    \centering
    \includegraphics[width=0.9\linewidth]{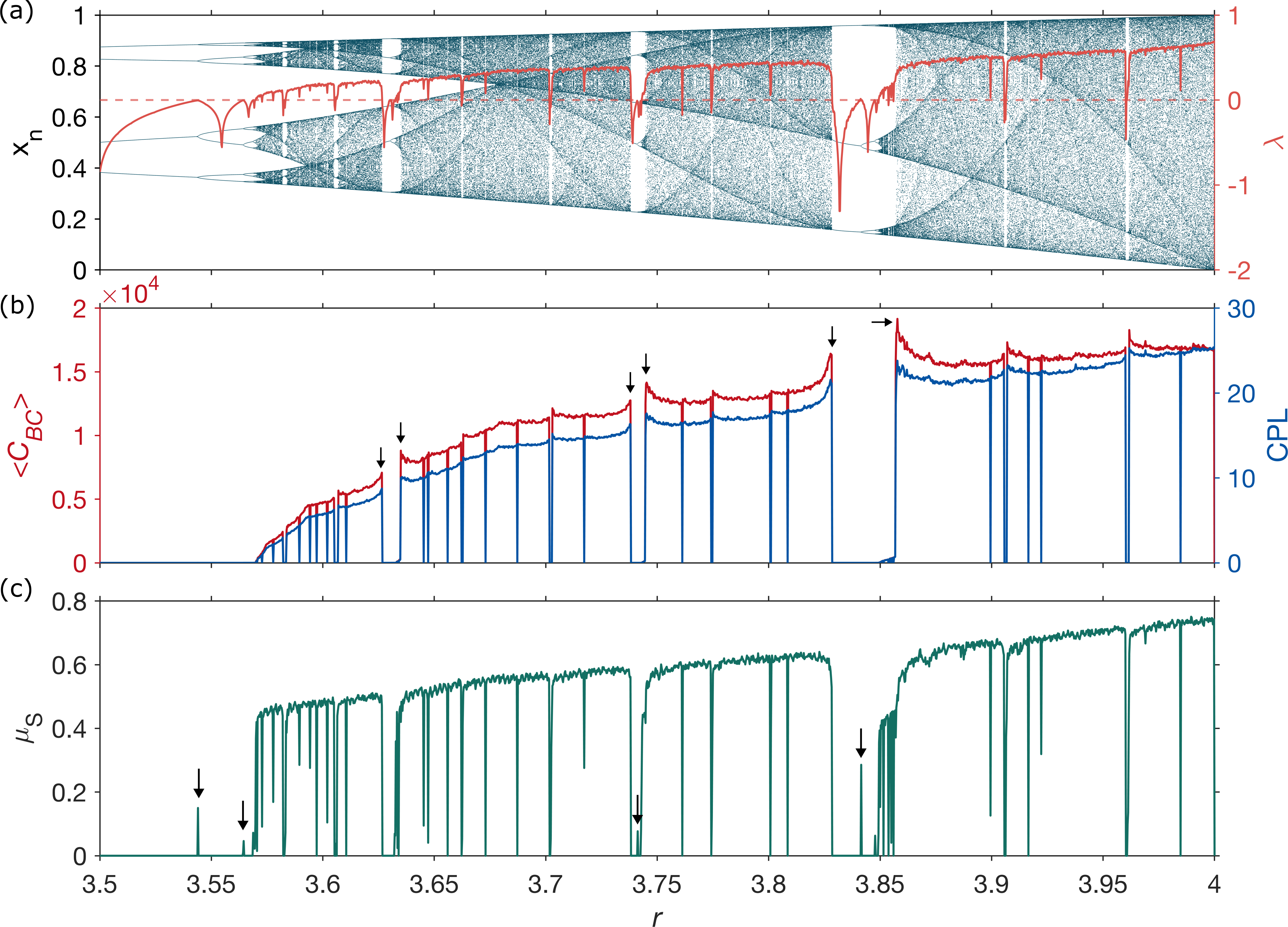}
    \caption{(a) Orbit diagram and the Lyapunov exponent $\lambda$ of the map (red curve; computed using) for $r$ varying from 3.5 to 4. (b,c) The variation of the average betweenness centrality $\langle C_{BC} \rangle$, the characteristic path length (CPL), and the average network entropy ($\mu_S$) with $r$. All the measures distinguish windows of periodicity and chaos. $\langle C_{BC} \rangle$ and CPL exhibits a spike at crisis event, depicted by an arrow in (b). Further, $\mu_S$ correlates with $\lambda$ and spikes at points of period doubling, as depicted by arrows in (c).}
    \label{fig:fig4_global_measures}
\end{figure*}
Destruction or widening of a chaotic attractor is accompanied by the occurrence of crisis-induced intermittency \cite{Grebogi1987CriticalIntermittency,Everson1987ScalingBoundary,Gu1984CrisesMaps,Kitano1984Symmetry-recoveringBistability}. Typically, intermittency is referred to as switching between chaotic and almost periodic dynamics at random intervals. In case of crisis-induced intermittency, the system switches between segments of two different types of chaos (Fig.~\ref{fig:orbit_supertrack}(b,c)). 
The logistic map exhibits `reverse bifurcation' where $2^n$ chaotic bands merge to give $2^{n-1}$ bands at $r_{n}^{b}$. These specific control parameter values are referred to as Misiurewicz points \cite{Pastor1996OnMaps,Pastor2001MisiurewiczMaps}. This feature of reverse bifurcation from $2^n$ bands to $2^{n-1}$ bands can be observed when the unstable fixed point born during the period doubling from $2^{n-1}$ to $2^n$ cycle collides with the attractor of $2^n$ band. Thus, there is a Misiurewicz point corresponding to every period doubling of the map.

To characterize the pattern of fluctuations and transitions exhibited by the logistic map, we encode the sequence generated at a particular control parameter $r$ onto a complex network topology. A network is a set of nodes connected by links representing a relation between the nodes. Time series data can be mapped onto network topology using various techniques \cite{Zou2019ComplexAnalysis}. Patterns across different scales in the time series can be extracted using network measures that capture local and global features of the network topology. 

We use amplitude transition networks to encode a sequence $x(t) = \{x_{1}, x_{2}, x_{3},\ldots, x_{n} \}$ as a network \cite{Shirazi2009MappingNetworks}. First, we discretize the range of amplitudes of the sequence of the map uniformly into bins; these bins constitute the nodes of the network. A link is established between node $i$ and $j$ if there exists a transition between the $i^{th}$ and $j^{th}$ bin and the weight of the edge is equal to the Markov chain transition probability between those two nodes. Hence, the elements of the connectivity (adjacency) matrix \textbf{A} of the network is given by $a_{ij} = {n_{i \rightarrow j}}/{\sum_k n_{i \rightarrow k}}$ (where $n_{i \rightarrow j}$ denotes the total number of transitions from node $i$ to $j$); i.e., $a_{ij} = p[x(t) \mid x(t - 1)]$, resulting in a weighted directed network. 

We apply the following type of network measures to detect the transitions: 
\textbf{(i)} \textit{Quasi-distance-based measures} are used such as mean average betweenness centrality ($\langle C_{BC} \rangle$), characteristic path length (CPL), and average network entropy ($\mu_S$) to characterize the global network topology. The betweenness centrality ($C_{BC}$) is a node-wise measure that quantifies how often a node lies in the shortest path between all pairs of nodes in the network \cite{Rubinov2009BrainDatasets.,Freeman1977ABetweenness}. The average betweenness centrality is given by $\langle C_{BC} \rangle =  1/N~\sum_{v = 1}^N g(v)$, where $g_v$ is the fraction of shortest paths from node $i$ to $j$ through node $v$ and $N$ is the number of nodes in the network. The measure is calculated for directed and weighted network using Dijkstra's algorithm. The characteristic path length (CPL) of a network quantifies the mean number of steps needed to reach one node from another in the network \cite{Rubinov2009BrainDatasets.,Albert2002StatisticalNetworks}. The average network entropy is given by $\mu_S = -\sum_{j} \sum_{i} a_{ij} \ln{a_{ij}}/{N}$, where $N$ is the number of nodes of the network \cite{Shannon1948ACommunication,Small2013ComplexDynamics}. The Shannon entropy of the edge weights quantifies the heterogeneity of interaction strengths in the network, i.e., $\mu_S$ would be higher for an amplitude transition network when the uncertainty and number of transitions between bins are higher and vice-versa \cite{Balaji2025NonmonotonicSystems}. 

\textbf{(ii)} \textit{Local network measures} are used to identify the structure of dynamical transitions in the orbit diagram. Chaotic trajectories have an infinite number of periodic orbits of varying stability in the phase space. For a one-dimensional map, these orbits translate to periodic points, and a chaotic trajectory switches from the vicinity of relatively unstable periodic point (UPP) to another \cite{Pawelzik1991UnstablePrediction, Auerbach1987ExploringOrbits}. The number of times the trajectory visits the vicinity of a periodic point depends on the relative stability of this point. We consider a series $x$ of the logistic map of length 20,000 after ignoring initial transients up to 1000 iterations for each control parameter $r$. The parameter $r$ is varied from 3.5 to 4 in steps of $\Delta r = 2.5 \times 10^{-4}$. We construct a network from $x$ at each $r$ by binning the amplitude from 0 to 1 into 2000 uniformly spaced bins. Thus, the network encodes the amplitudes attained as well as the frequency of transition between distinct amplitude scales in a trajectory. 

We find that both $\langle C_{BC} \rangle$ and CPL exhibit an increasing trend with $r$, correlating with the increasing size of the attractor (Fig.~\ref{fig:fig4_global_measures}(b)). Also, these measures attain low values for periodic dynamics but large values during chaotic regimes (Fig.~\ref{fig:fig4_global_measures}(b)). For a network constructed at a particular $r$, $\langle C_{BC} \rangle$ reflects the non-uniformity in amplitude jumps in the iterative sequence of $x_n$. The network constructed from any period-$k$ orbit is cyclic and regular. Thus, the $C_{BC}$ of all nodes is almost equal and low, resulting in a low value of $\langle C_{BC} \rangle$. On the other hand, a chaotic sequence implies jumps between different amplitude levels in a disorderly manner and reflects a disordered network topology. The chaotic orbits thus have a non-uniform distribution across amplitudes, leading to a higher average betweenness centrality and average path length. 
Interestingly, both $\langle C_{BC} \rangle$ and CPL exhibit spikes at $r = 3.635, 3.745, 3.857, 3.961$ (depicted by arrows in Fig.~\ref{fig:fig4_global_measures}(b)), which coincide with the locations of chaotic attractor crisis. The intermittent switching between two chaotic regimes of the attractor during an interior crisis induces a much higher $C_{BC}$ for the intermediate nodes (amplitude bins), thus increasing the average betweenness centrality of the network. Furthermore, the sudden increase in the attractor size requires more steps to transverse from one node to another, leading to an increase in CPL of the network.

The variation of $\mu_S$ with $r$ (Fig.~\ref{fig:fig4_global_measures}(c)) is consistent with the Lyapunov exponent $\lambda$ of the map, i.e., $\mu_S$ approaches zero for periodic dynamics and non-zero values during chaotic dynamics. $\mu_S$ also exhibits a spike at $r$ corresponding to every period doubling of the map (indicated by $\downarrow$ in Fig.~\ref{fig:fig4_global_measures}(c)). Also, note that all network measures appear to fluctuate in a certain range of control parameters (see for e.g., around $r \approx$ 3.85). These fluctuations occur owing to the presence of periodic windows within the chaotic bands within a very narrow range of $r$.

\begin{figure*}
    \centering
    \includegraphics[width=\linewidth]{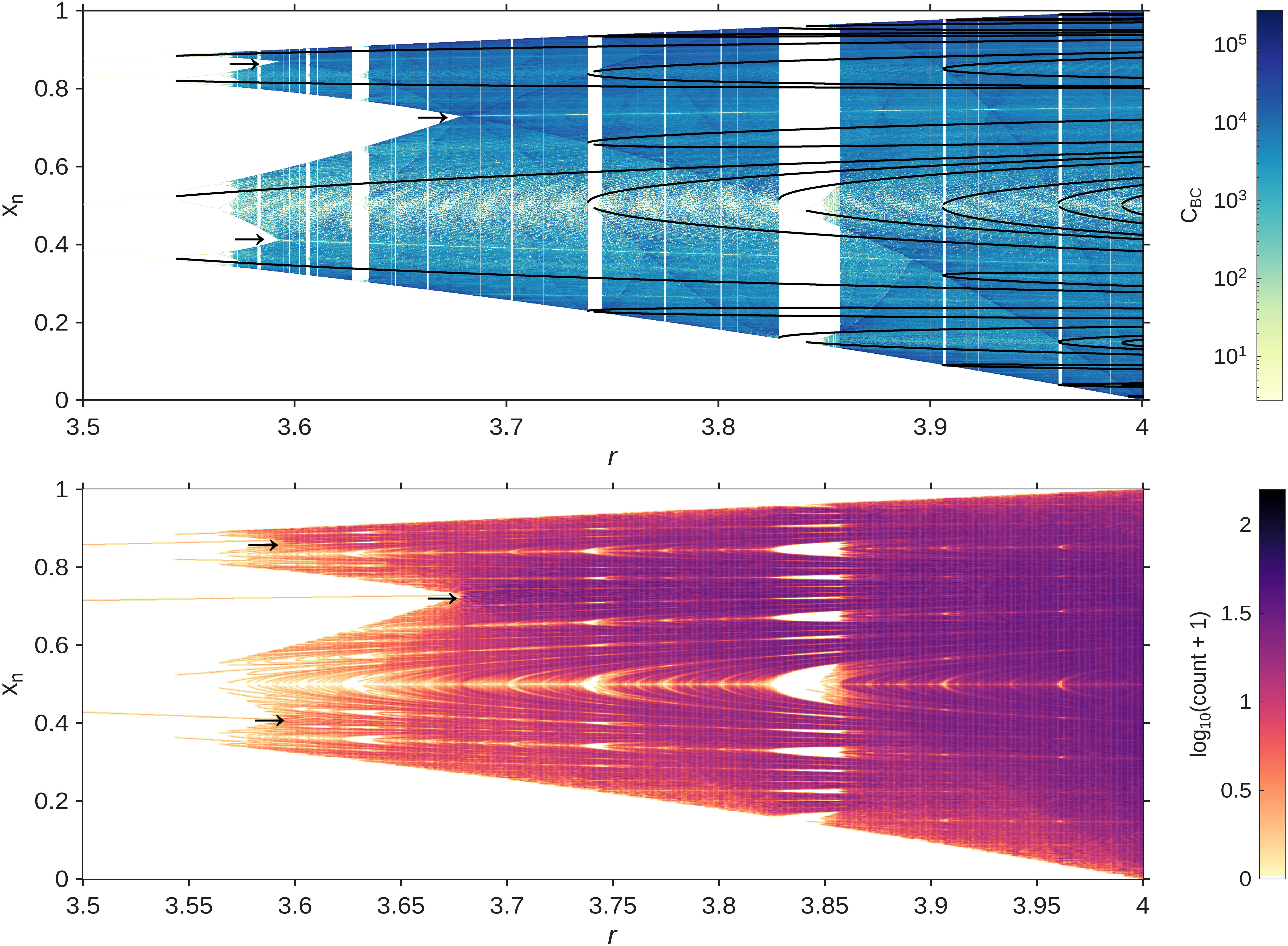}
    \caption{(a) Node-wise distribution of betweenness centrality $C_{BC}$ over $r\in[3.5,4]$ and unstable periodic points (UPP) of periods $3$ to $6$ are indicated in black. High $C_{BC}$ ridges align with supertrack functions, while low $C_{BC}$ bands correspond to UPPs. (b) Density of unstable periodic orbits up to period $p_{\max}=20$ in the $(r,x)$ plane, obtained from solutions of $F^p(x)=x$ filtered for exact period and instability ($|\lambda|>1$). The density is computed by binning $(r,x)$ into a $700\times700$ grid and visualized as $\log_{10}(\mathrm{count}+1)$ to enhance contrast across orders of magnitude. Node-wise distribution of betweenness centrality $C_{BC}$ for $r$ from 3.5 to 4. The distribution highlights the supertrack functions and unstable fixed points of the map as streaks of high and low $C_{BC}$, respectively.
    }
    \label{fig:bwt_local}
\end{figure*}

Figure~\ref{fig:bwt_local}(a) shows the variation of node-wise distribution of betweenness centrality ($C_{BC}$) for $r$ varied from $3.5$ to $4$. Note that $C_{BC}$ quantifies how often a node lies in the shortest path between all pairs of nodes in the network.  Since, $C_{BC}$ of a node captures information about the paths that pass through it, this measure is a quasi-local measure that encodes both local properties and global connectivity of the network. We observe low values of $C_{BC}$ for some nodes in Fig.~\ref{fig:bwt_local}(a) indicating that only a few paths traverse through these nodes when the trajectory jumps from one amplitude bin to another. Evidently, the trajectory of the map at a given $r$ value rarely evolves in the vicinity of the periodic points in the state space that are relatively more unstable. \textcolor{black}{Thus, a lower value of $C_{BC}$ corresponds to amplitude bins that are rarely traversed along shortest path, denoting repelling regions of phase space associated with UPPs. Conversely, nodes with high $C_{BC}$ lie on the paths connecting many pairs of nodes, indicating that an orbit is preferential through these bins, coinciding with the attracting character of these regions they occupy. This is further confirmed by plotting UPPs of periods $p\in[3,6]$, shown via black points superimposed on the orbit diagram in Fig.~\ref{fig:bwt_local}(a), that coincide with the $\subset$-shaped locus formed by points of low $C_{BC}$.}

We observe a hierarchy of such $\subset$-shaped locus of amplitude bins characterized by low $C_{BC}$. We conjecture that this family of $\subset$-shaped curves depicts the relative stability of points. To confirm this, we visualize the distribution of several UPPs as shown in Fig.~\ref{fig:bwt_local}(b). UPPs are computed by finding the fixed points of the $p^{\text{th}}$ iterate \cite{SANDER2012} of the map upto $p=20$ and visualized via heatmap of their density in discrete amplitude-bins. 
A comparison between the $C_{BC}$ plot in Fig.~\ref{fig:bwt_local}(a) and the density of UPPs in Fig.~\ref{fig:bwt_local}(b) confirms that $C_{BC}$ effectively delineates the skeleton structure formed by UPPs of the map. \textcolor{black}{Stable and unstable periodic points together form the skeleton of the phase space, governing the path the trajectories follow as they evolve in phase space. The network reconstructs this skeleton from time-series data alone, without requiring the analytical knowledge of the fixed points of the map.} We also find the emergent family of curves with high values of $C_{BC}$ in Fig.~\ref{fig:bwt_local}(a), that clearly correspond to the supertrack functions (`$Q$'-curves \cite{ROSS2009TheToo}). \textcolor{black}{This finding reiterates the well known fact that orbits comprising points on or close to the supertrack function have relatively higher density of trajectory points; these high $C_{BC}$ ridges effectively bundle the paths of the orbit through a concentrated set of amplitude bin, making them the attracting counterparts to the repelling UPP locus. }In summary, the variation of $C_{BC}$ unravels both, the attracting and the repelling curves of the map and exhibits the basic signature of the \textit{bifurcation diagram} of the logistic map. 

\begin{figure*}
    \centering
    \includegraphics[width=\linewidth]{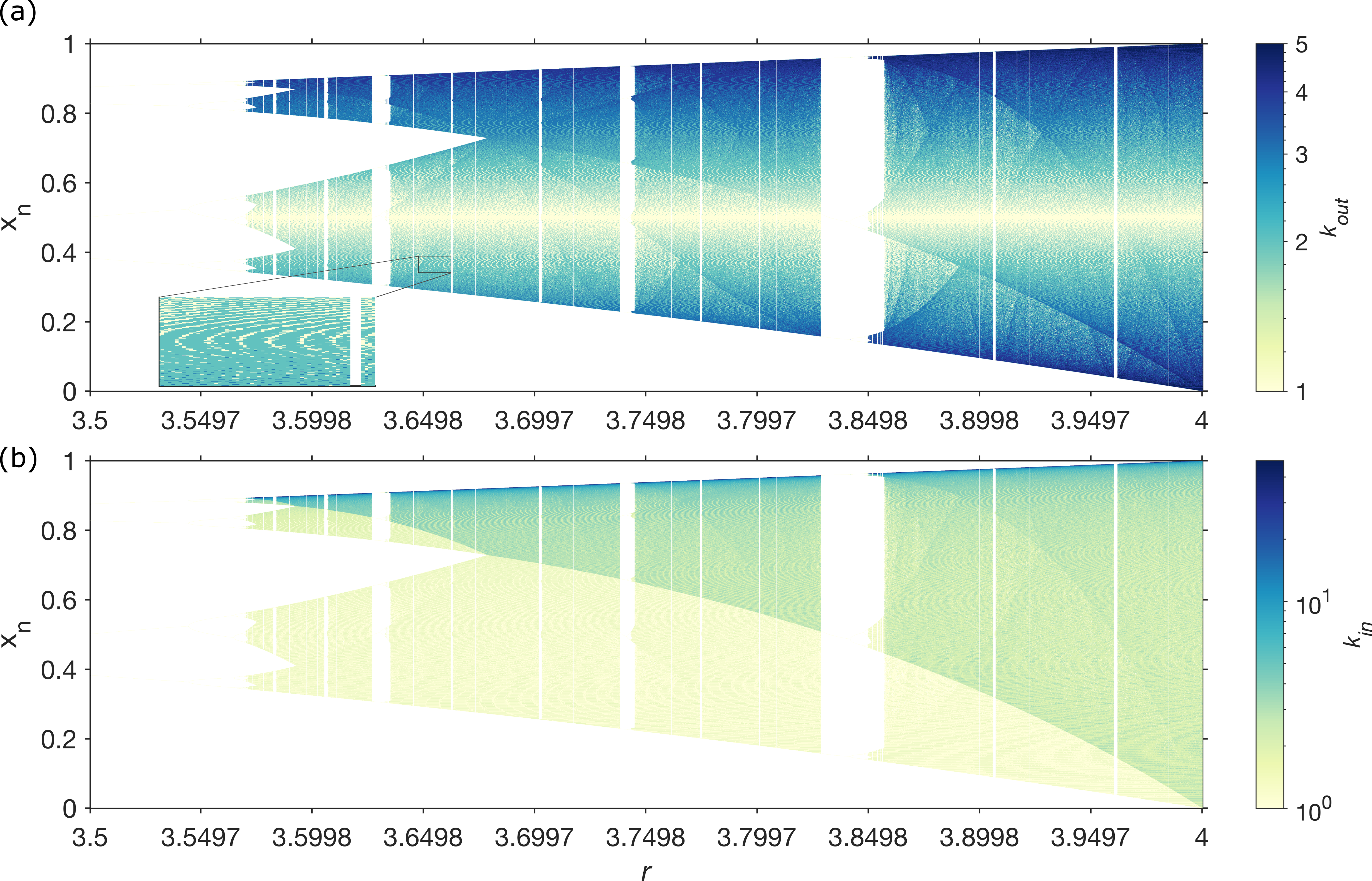}
    \caption{Node-wise distribution of (a) out-degree $k_{out}$ and (b) in-degree $k_{in}$ centrality for $r$ varied form 3.5 to 4. $k_{out}$ denotes the number of unique outgoing links from a specific bin, which could be viewed as the tendency of the unstable periodic orbit to stay near the vicinity of an amplitude bin, delineating the relative stability of the amplitude bin.}
    \label{fig:degree_local}
\end{figure*}
Next, we compute the in-degree $k_{in}$ and the out-degree $k_{out}$ as the number of unique links directed to and from a node in the network. Mathematically, we can express the in- and out-degree of a node $j$ using the Heaviside function $\theta(x)$ as $k_{in}|_j = \sum_{i} \theta(\mathrm{\textbf{A}}_{ij})$, and $k_{out}|_j = \sum_{i} \theta(\mathrm{\textbf{A}}_{ji})$. Thus, the degree measure denotes the number of transitions occurring into (or out of) an amplitude bin. We find firstly, that the supertrack functions are evident in the node-wise $k_{in}$ and $k_{out}$ distributions (Fig.~\ref{fig:degree_local}). The supertrack functions are attracting curves thus leading to frequent transitions to and from these regions and consequently resulting in higher in- and out-degrees. 
Further, we can see that $k_{out}$ and $k_{in}$ highlights an asymmetry and non-uniformity in the orbit diagram of the logistic map (Fig.~\ref{fig:degree_local}). 
Specifically, we find that the lower half region exhibits lower in-degrees but comparable out-degrees compared to the upper-half region of the map with greater amplitudes. Interestingly, the boundary which separates the regions of significant in-degree variations can be clearly identified as a combination of certain supertrack functions (Fig.~\ref{fig:degree_local}(b)). On zooming in, we also find small $\subset$-shaped curves formed by points of relatively similar degrees highlighting the non-uniformity in the distribution of the out-degree (shown in the inset of Fig.~\ref{fig:degree_local}(a)), suggesting an underlying structure in the way transitions are organized across the state space. 
\begin{figure*}
    \centering
    \includegraphics[width=\linewidth]{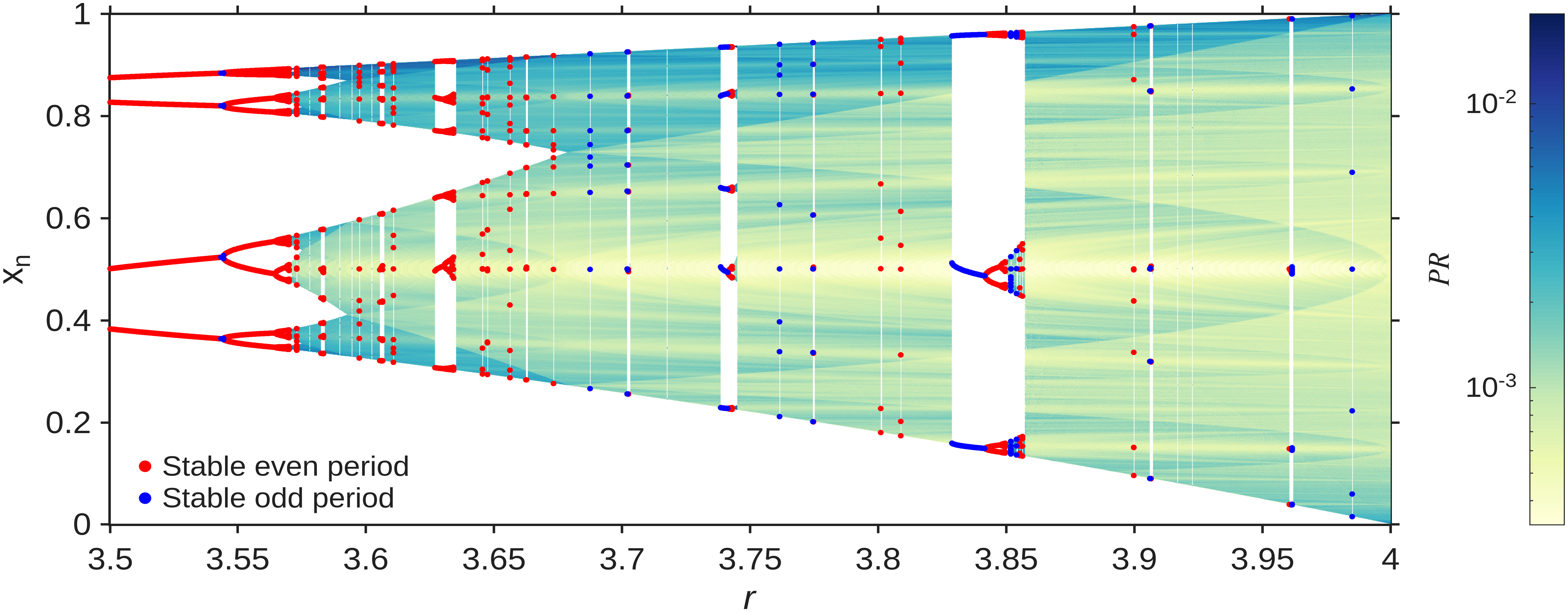}
    \caption{Node-wise distribution of PageRank centrality for $r$ varied form 3.5 to 4. The local network measure displays perplexing inverted $\subset$-structures on the orbit diagram, intersecting at the Misiurewicz point.}
    \label{fig:pg_local}
\end{figure*}

Finally, we compute the node-wise PageRank centrality ($PR$) derived from the network constructed at each $r\in[3.5,4]$ (Fig.~\ref{fig:pg_local}). PageRank quantifies the relative importance of a node in a network, i.e., a random walker on a directed network is more likely to end up on a node with high $PR$ \cite{Rubinov2009BrainDatasets.}. Unlike the distributions of $C_{BC}$ and $k_{out}$, supertrack functions are not the prominent feature captured in the distribution of $PR$. Instead, we identify the locus of points of high $PR$ arranged along diverging curves that emerge from various Misiurewicz points. Moreover, these diverging curves appear to be inverted $\subset$-shapes; for example, see the $\subset$-shape with the vertex starting at $r\approx4$, and $x_n\approx0.5$. Interestingly, these curves of high $PR$ separate regions of the orbit diagram that comprise several SPPs from regions that are sparsely populated by SPPs. This is evident from the fact that most SPPs lie along the axis of one of inverted $\subset$-shaped curves evident in the distribution of $PR$; for example see all the SPPs that occur at $x_n\approx 0.5$.

In summary, using complex networks we decipher the pattern of temporal fluctuations in the logistic map. The global and quasi- local path-based measures derived from the constructed network reveal different dynamical regimes and transitions by characterizing the amplitude transitions encoded in the network topology. Also, the information theoretic measure correlate very well with the Lyapunov exponent, making it an alternative and robust measure to characterize the chaotic nature of dynamical systems. Importantly, local centrality measures, such as $C_{BC}$ and degree, reveal the skeleton of the orbit diagram formed by stable and unstable periodic points and supertrack functions. Further, the PageRank reveals novel inverted $\subset$-shaped boundaries that intersect at Misiurewicz points and separates regions containing stable periodic points. Stable and unstable periodic points guide orbit evolution in phase space. The network-based measures, such as $C_{BC}$, degree centrality, and PageRank, effectively resolve the skeleton formed by the SPPs and UPPs of a map exhibiting intricate dynamical transitions. The network reconstructs the skeleton of the orbit diagram without explicit knowledge of the analytical fixed points. These findings suggest that the network constructed from complexity of fluctuations in a system can reconstruct the invariant dynamical structures of the phase space directly from observations remarkably well.
\footnote{See supplementary material for (i) a systematic analysis showing that the network measures are robust to small variations in the number of bins and sequence length, (ii) effect of noise on the network entropy $\mu_S$.} It will be an interesting direction for future works to corroborate these findings for high-dimensional maps as well as flows, and to detect invariant structures in the attractor. 

\begin{acknowledgments}
Acknowledgments: This research was supported by IoE-IITM Research Initiatives (SP/22-23/1222/CPETW
OCTSHOC) grant provided to RIS. ST acknowledges the Prime Minister's Research Fellowship (PMRF) from Govt. of India. \\~\\
\end{acknowledgments}
\vspace{-25pt}
\noindent {Data availability:} Data sharing is not applicable to this article as no new datasets were generated. All the information necessary to reproduce the results is provided in the manuscript.

\noindent {Author contributions:} Conception: ST, NM, JK, RIS; Methodology: AB, ST, SV, NM; Analysis and visualization: AB, SV; Interpretation: AB, ST, SV, NM, JK, RIS; Supervision: NM, JK, RIS; Writing: AB, ST, SV; Review \& editing: NM, JK, RIS

\bibliography{references}

\end{document}